\documentclass[journal]{IEEEtran}
\usepackage{graphicx}
\ifCLASSINFOpdf
 
\else

\fi

\begin{document}

\title{
  Implementation of Memristor in Bessel filter with RLC components\\
  }

\author{Galymzhan~Torebayev and Anju Nandakumar\\
   
\IEEEauthorblockA{ Electrical and Computer Engineering, Nazarbayev University, Astana, Kazakhstan
galymzhan.torebayev@nu.edu.kz, anju.nandakumar@nu.edu.kz}
}

\maketitle

\IEEEpeerreviewmaketitle
\begin{abstract}
The Bessel filters are optimized to collect competent transient response due to a linear phase in the passband. In other words, during the filtering process, there will be comparatively impoverished frequency response with lower amplitude inequity. Memristor is asserted as a passive, two-terminal essential component of the circuit and the use of such element in schemes as an adjustable resistance allows the realization of memory-resistor based analog circuits, which achieve the wide range of specific parameters. The application of RLC circuit for Bessel filter prototype is theoretically expected to behave in a positive way, however, the further simulations with software and analysis of the results will reveal the nature of the effect.   
\end{abstract}

\begin{IEEEkeywords}
Analog Circuits Design, Memristor, Bessel filter, memristor, RLC circuits analog filter
design, memristor-based filter.
\end{IEEEkeywords}

\section{Introduction}

\IEEEPARstart {}
The filters in real life are an inevitable component of each and every electronic device with modern technologies. Environmental noise is not desired in device performance and therefore must be filtered out. 
This project proposes the low pass filter with Bessel filter arrangement using memristor and RLC components. The second order of Bessel filter is used for simplicity in achieving a variety of results by manipulating the characteristics of the components. Increasing the order of Bessel filter leads to enhancement in performance as removing the ringing and overshooting effects from the signal \cite{Payana}. The Bessel filter is commonly studied for obtaining more favorable step-transient response \cite{Filanovsky}. It also provides the maximally flat response, which is best for audio frequencies, in addition to fast settling time \cite{Susan}.
Memristor means memory resistor. It remembers the voltage input that passed through it and depending on the present voltage magnitude remains active or inactive. The voltage can pass in both ways in a resistor, whereas in memristor it flows only in one direction. Therefore, there is decrease in a leakage voltage in the circuit.

\section{Components and Description}

Regardless of whether the frequency is low or high, it is fundamental that resistors obey the Ohm’s law, which is V=I*R. Nevertheless, in the real instances, non-linear changes occur to resistors, because of impacting factors as temperature due to heat dissipation or use of semiconductor with saturation effect. 
\newline
Capacitors are one of the essential components in electrical circuits that typically made of a couple of electrical conductors that capture energy. This energy occurs in the field between the conducting plates of the capacitor. Capacitors can be specified by the amount of energy and charge that the conductors are able to store inside at constant voltage. The quantity of energy that can be stored identifies the main parameter of the capacitor, which is capacitance. In addition, the physical geometry of capacitor contributes to the capacitance value. For the capacitors Q=CV. The charge is directly proportional to voltage and capacitance.
\newline
Memristor stands for “memory resistor”. In electrical circuits, the memristor is two terminal passive element that can replace the resistor in theoretical approach.  The study shows that the behavior of the memristor is similar to the one of synapses between neurons in human brain. Memristor “memorizes” the voltage that passed through most recently and remains still after supply is turned off. The primary parameter for the memristor is its memristance. The memristance illustrates the changing rate of magnetic flux with charge dependence with equation M=d $\Phi$ /dq. The current experienced by memristor in the past affects the memristance of the given element. \cite{Ascoli}.
\newline 
The list of components then proceeds to the operational amplifier. The operational amplifier is among the list of the most widely used devices in analog circuits. Op-amp is a direct current coupled voltage amplifier with significantly high gain value. The amplifier can perform in either differential, integral or even low pass filter mode (fig 1). The input impedance, as well as the common mode rejection rate and voltage gain of an open loop, can be infinite.  Additionally, the input offset voltage and zero output impedances are important features that correspond to the operational amplifier. \cite{Gao}.

\begin{figure}  [!ht] 
\begin{center}  
\includegraphics [width=8.5cm]{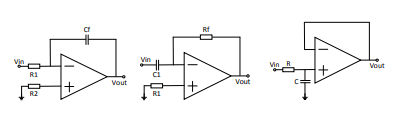} 
\caption{\small \sl The Operational Amplifier modes}
\end{center}     
\end{figure}

\section{Simulation}
In the simulation part of the project initially, the characteristics of the memristor were tested. Figure 2 illustrates the circuit built in LTspice and Figure 3 shows the simulation results of the memristor model in Matlab. 
\newline 
The next step in the project was testing the performance of the Bessel filter of 2nd order with 3 resistors, no memristor added yet (fig 4). The supply component adjusted to send AC signal with starting frequency of 100 Hz and stop frequency of 1MEG Hz. Figures 5 and 6 depict the Matlab simulation which is obtained by importing the graph data from LTSpice to Matlab software. Further graphs are obtained in the same way. Afterward, as it was supposed theoretically, the resistors are replaced by memristors in figure 7 and the results were quite similar (figures 8.9).  Then the circuit was decided to leave one resistor and have two memristors as the RLC elements are being studied as well (fig 10). It is noticeable that amplitude response has the same shape and small changes (fig 11). On the other hand, the phase response has a significant difference, the graph stars with -180 degrees at the start and then follows the same path of phase response in Bessel filter with 3 memristors (fig 12). The manipulations with parameters of the components show that the graph of amplitude and phase response depends strongly on the magnitude of the remaining resistor. The figures 13 and 14  shows that phase and amplitude responses change totally when the resistor has a value of 100 Ohm instead of 15 kOhm. 

\begin{figure}  [!ht] 
\begin{center}  
\includegraphics[width=8.5cm]{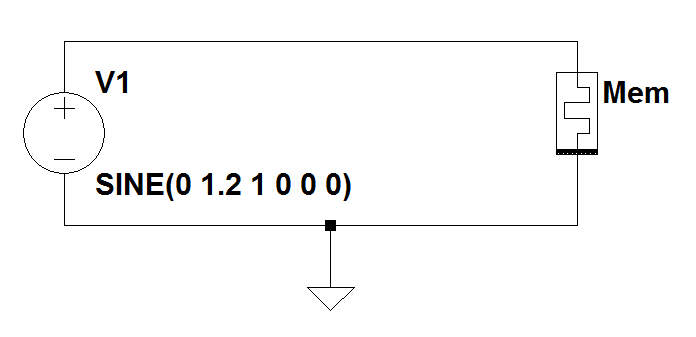} 
\caption{\small \sl The memristor Model in LTspice}
\end{center}     
\end{figure}

\begin{figure}  [!ht] 
\begin{center}  
\includegraphics[scale=0.4]{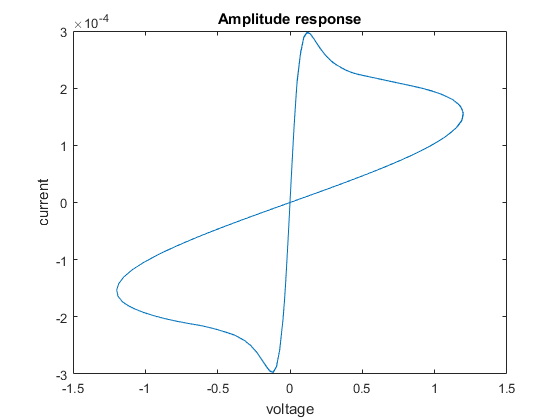} 
\caption{\small \sl The IV characteristics of memristor (hysteresis)}
\end{center}     
\end{figure}

\begin{figure}  [!ht] 
\begin{center}  
\includegraphics[width=8.5cm]{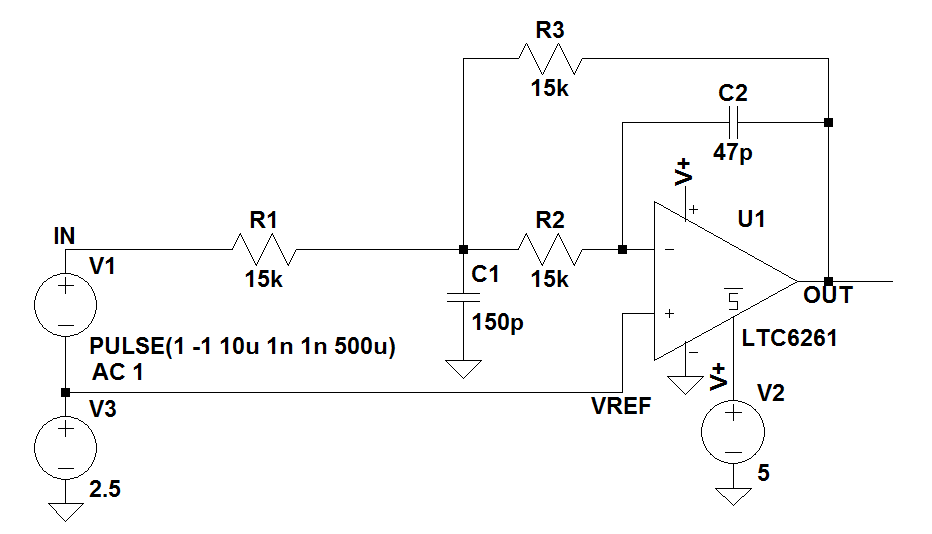} 
\caption{\small \sl The 2nd order Bessel filter circuit}
\end{center}     
\end{figure}

\begin{figure}  [!ht] 
\begin{center}  
\includegraphics[scale=0.4]{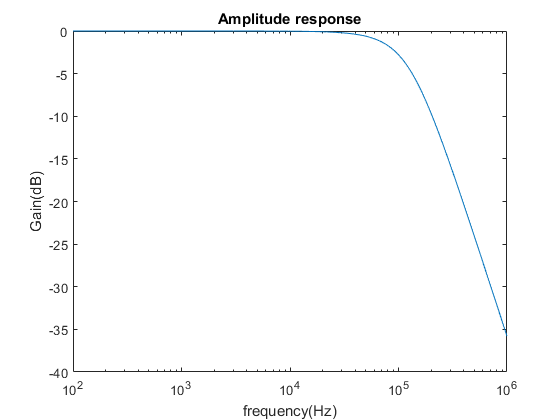} 
\caption{\small \sl The amplitude response of 2nd order Bessel filter}
\end{center}     
\end{figure}

\begin{figure}  [!ht] 
\begin{center}  
\includegraphics[scale=0.4]{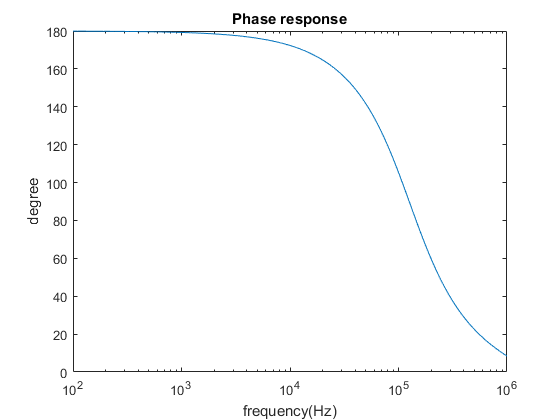} 
\caption{\small \sl The phase response of 2nd order Bessel filter}
\end{center}     
\end{figure}

\begin{figure}  [!ht] 
\begin{center}  
\includegraphics[width=8.5cm]{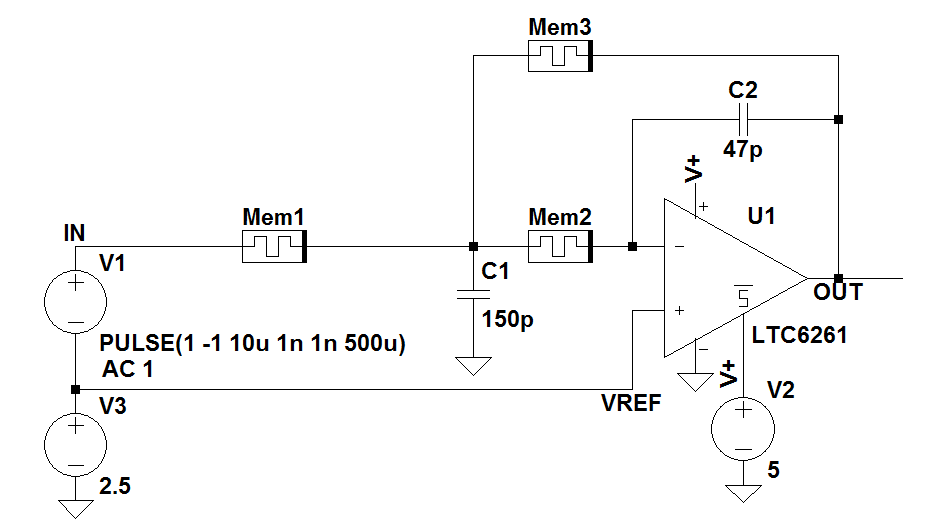} 
\caption{\small \sl The Bessel filter with memristors}
\end{center}     
\end{figure}

\begin{figure}  [!ht] 
\begin{center}  
\includegraphics[scale=0.4]{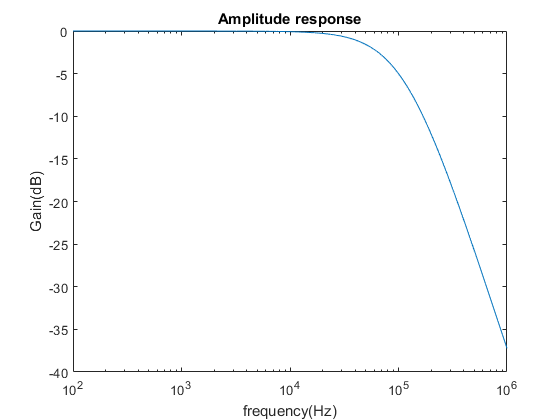} 
\caption{\small \sl The amplitude response of Bessel filter with memristors}
\end{center}     
\end{figure}

\begin{figure}  [!ht] 
\begin{center}  
\includegraphics[scale=0.4]{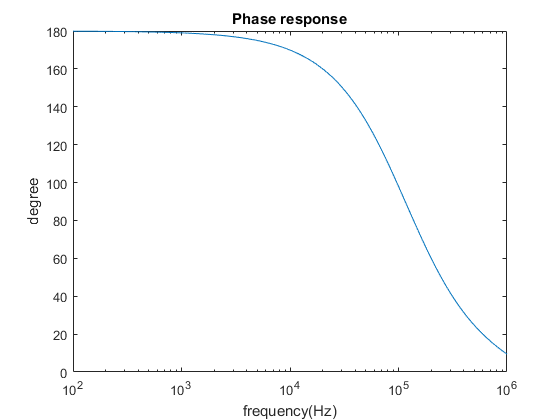} 
\caption{\small \sl The phase response of Bessel filter with memristors}
\end{center}     
\end{figure}

\begin{figure}  [!ht] 
\begin{center}  
\includegraphics[width=8.5cm]{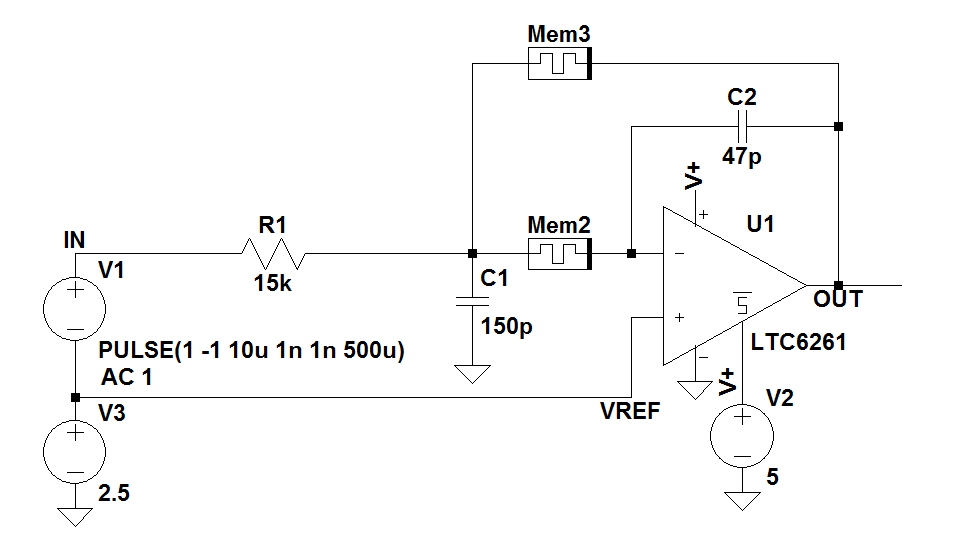} 
\caption{\small \sl The Bessel filter with 1 resistor, 2 memristors}
\end{center}     
\end{figure}

\begin{figure}  [!ht] 
\begin{center}  
\includegraphics[scale=0.4]{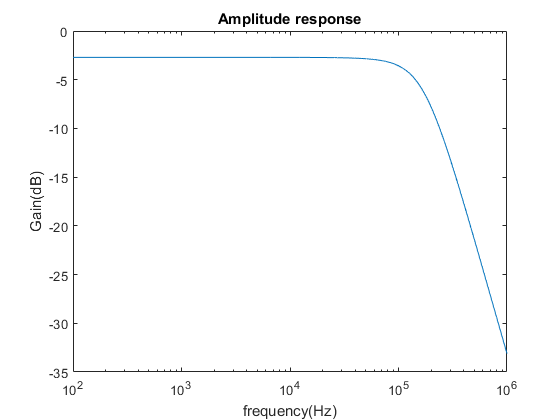} 
\caption{\small \sl The amplitude response of circuit in figure 10}
\end{center}     
\end{figure}

\begin{figure}  [!ht] 
\begin{center}  
\includegraphics[scale=0.4]{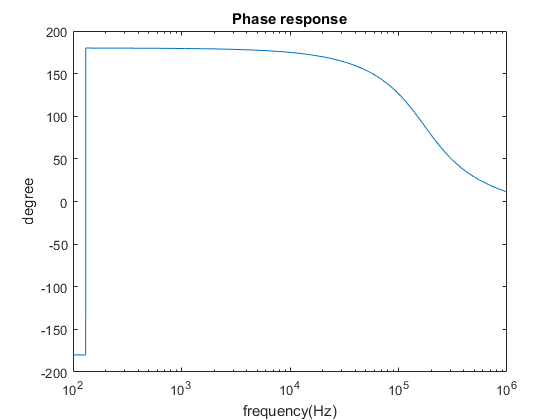} 
\caption{\small \sl The phase response of circuit in figure 10}
\end{center}     
\end{figure}

\begin{figure}  [!ht] 
\begin{center}  
\includegraphics[scale=0.4]{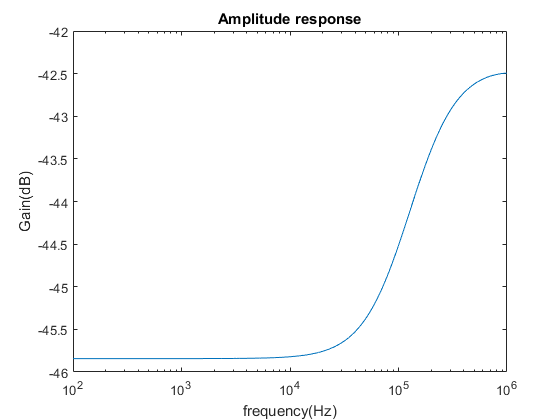} 
\caption{\small \sl The amplitude response of circuit in figure 10 with 100ohm resistor}
\end{center}     
\end{figure}

\begin{figure}  [!ht] 
\begin{center}  
\includegraphics[scale=0.4]{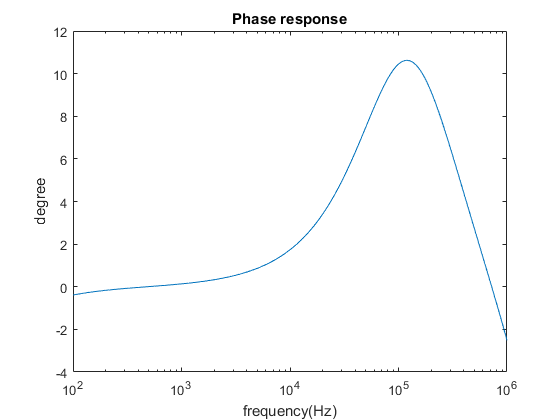} 
\caption{\small \sl The phase response of circuit in figure 10 with 100ohm resistor}
\end{center}     
\end{figure}

\section{ALTERNATIVE SIMULATIONS}
The LTSpice is outstanding for circuit building and analysis. However, there are other software to use for such purposes. It is important to mention such tools as PSpice and Matlab. The following specifications depict the simulation of Bessel filter via Matlab software.
\newline
\textbf{Limitations:} Low pass filters with Bessel arrangement have deteriorating magnitude response, which is also true for low pass Butterworth filters. The distinguishing feature of Bessel, when comparing with Chebyshev or Butterworth filters, is the most gradual roll-off and the requirement for the highest order filter, to meet the specifications regarding the attenuation.The state space form is maximally authentic at high order filtering, next precise form is zero, pole, gain form. On the other hand, the transfer function coefficient form is the last in the list of accuracy. Further, coding the memristor prototypes for Matlab is hard to achieve because of the complexity.
\newline
\textbf{Algorithms:} In Matlab software Bessel functions have four-stepped workflow as follows:
\newline	 
Firstly, low-pass analog model  gain, pole and zero points are to be found by using special function \textit{besselap}. Secondly, the state space form is created by converting the pole, zero points and gain. Thirdly, the low-pass model is reconstructed into a low-pass filter with desirable specifications. Eventually, as it is vital, state space filter is transformed back to pole-zero transfer function form
\newline
\textbf{Description:} In Matlab software, the function [z,p,k]=besselap(n) recovers the pole-zero and gain of an nth order Bessel low-pass filter model. In the equation, “p” stands for poles column, “k” is the gain, and z is an empty matrix, as there are no zero points. In addition, the magnitude of variable “n” cannot exceed 25. The transfer function to analyze the Bessel filter is: 
\newline
\(H(s)=/((s-p(1))*(s-p(2))*(s-p(3))*(s-p(4))*⋯*(s-p(n)))\)
The purpose of “besselap” function is to apply normalization on the poles and gain, to achieve desirable behavior of Bessel function at different ranges of frequencies. In addition, the same order Bessel and Butterworth prototypes must have asymptotically equal poles and gain. At the unity cutoff frequency, the value of the filter is no more than 0.707, which is the square root of ½. In analog circuits, the Bessel filters are specified by a group delay parameter. This parameter is described to be maximally flat at frequency=0 and approximately constant along the passband. When the frequency is equal to 0, the group delay equation is:
\(((2n)!/(2^n*n! ))^(1/n)\).

\begin{figure}  [!ht] 
\begin{center}  
\includegraphics[width=8.5cm]{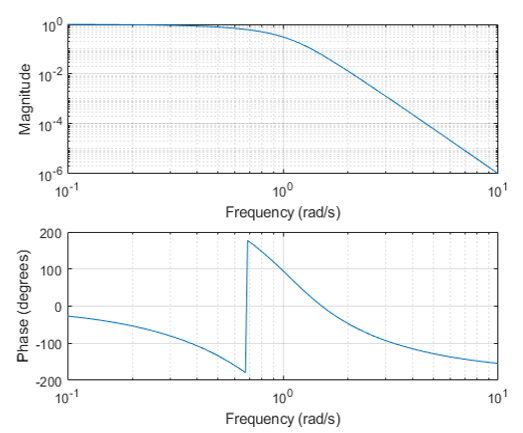} 
\caption{\small \sl Phase and Magnitude response}
\end{center}     
\end{figure}

\begin{figure}  [!ht] 
\begin{center}  
\includegraphics[scale=0.5]{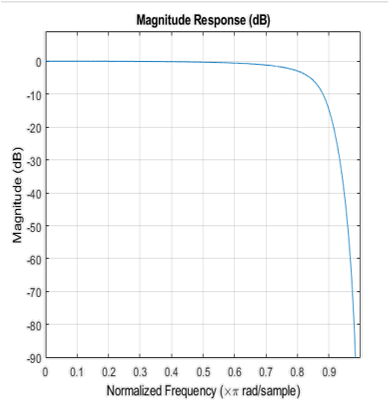} 
\caption{\small \sl Magnitude response with normalized frequency}
\end{center}     
\end{figure}

\section{Further research}
Firstly, results can be enhanced by considering the combination of Bessel filter with another filter called Butterworth for obtaining better characteristics as faster settling time, decreased overshooting effect, step-transient response and etc. Secondly, the inductive component can be added to the circuit in order to achieve greater performance result. In addition, the filter group delay could be used for examination of performance \cite{Yigang}.

\section{Conclusion}
A new version of Bessel filter design based on memristor with RLC elements is presented. The proposed circuit achieved the desired theoretical results with frequency response. The circuit is useful for filtering the high-frequency components, in other words, work as a convenient low pass filter. As the innovation to existing circuits proposed project utilizes memristor in performance instead of resistors. There are significant on-going researches to understand memristor dynamics, and physics-based models are being developed \cite{Secco}.

\ifCLASSOPTIONcaptionsoff
  \newpage
\fi


\begin{thebibliography}{1}

\bibitem{Payana}
M. Pavana, N. Shashikala and J. Manisha, "Design, development and Comparative performance analysis of Bessel and Butterworth filter for Nadi Pariksha yantra", 2016 IEEE International Conference on Engineering and Technology (ICETECH), 2016.
\bibitem{Filanovsky}
I.M. Filanovsky, "Bessel-Butterworth transitional filters", 2014 IEEE International Symposium on Circuits and Systems (ISCAS), 2014.
\bibitem{Susan}
D. Susan and S. Jayalalitha, "Bessel filters using simulated inductor", 2011 INTERNATIONAL CONFERENCE ON RECENT ADVANCEMENTS IN ELECTRICAL, ELECTRONICS AND CONTROL ENGINEERING, 2011.
\bibitem{Gao}
Fei Gao,Shengxi Diao,Di Zhang,Shuang Wang, "A Gm-C operational amplifier for analog signal processing", 2017 10th International Congress on Image and Signal Processing, BioMedical Engineering and Informatics (CISP-BMEI), 2017.
\bibitem{Ascoli}
A. Ascoli, R. Tetzlaff, F. Corinto, M. Mirchev and M. Gilli, "Memristor-based filtering applications", 2013 14th Latin American Test Workshop - LATW, 2013.
\bibitem{Secco}
Jacopo Secco,"Flux–Charge Memristor Model for Phase Change Memory", IEEE Transactions on Circuits and Systems II: Express Briefs, vol. 65, no. 1, pp. 111-114, 2018.
\bibitem{Yigang}
Yigang He, Jinguang Jiang and Yichuang Sun, "CMOS R-MOSFET-C fourth-order Bessel filter with accurate group delay", 2002 IEEE International Symposium on Circuits and Systems. Proceedings (Cat. No.02CH37353).
\bibitem{}
Maan Akshay Kumar, Deepthi Anirudhan Jayadevi, and Alex Pappachen James. "A survey of memristive threshold logic circuits." IEEE transactions on neural networks and learning systems 28.8 (2017): 1734-1746.

\end{thebibliography}
\end{document}